\documentstyle[fleqn,a4,pstricks,pst-plot]{espart}

\begin{document}

\begin{frontmatter}

\title{Extended Shell Model Calculation for even $N=82$ Isotones with a 
Realistic Effective Interaction}

\author{A.\ Holt, T.\ Engeland and E.\ Osnes}

\address{Department of Physics, University of Oslo, N-0316 Oslo, Norway}

\author{M.\ Hjorth-Jensen}

\address{Nordita, Blegdamsvej 17, DK-2100 K\o benhavn \O, Denmark}

\author{J.\ Suhonen}

\address{Department of Physics, University of Jyv\"{a}skyl\"{a},
              P.O.Box 35, FIN-40351 Jyv\"{a}skyl\"{a}, Finland}

\maketitle

\begin{abstract}
The shell model within the $2s1d0g_{7/2}0h_{11/2}$ shell is applied to 
calculate nuclear structure properties of the even $Z=52 - 62$, $N=82$ 
isotones. The results are compared with experimental data and with the 
results of a quasiparticle random-phase approximation (QRPA) calculation.
The interaction used in these calculations is a realistic 
two-body $G$-matrix
interaction derived from modern meson-exchange potential models for the
nucleon-nucleon interaction. 
For the shell model all the 
two-body matrix elements are renormalized by the $\hat{Q}$-box method 
whereas for the QRPA the effective interaction is defined 
by the $G$-matrix.
\end{abstract}

\end{frontmatter}

\section{Introduction}

In recent years the $N=82$ isotones have been a subject of great interest and 
experimental data for the isotones are rather well established.  These nuclei 
show a high degree of regularity, in the sense that there are many 
similarities between neighbouring even-even nuclei. Abbas \cite{a84} even 
goes as far as to claim that all the $N=82$ and $Z=58, 60, 62, 64, 66, 68, 70$
nuclei were doubly magic. At least for $^{146}$Gd the doubly magic character 
and the subshell closure at $Z=64$ is rather well established 
\cite{wsg89,yjk86}.

During the last few years experimental evidence for 
low-lying octupole-octupole
and octupole-quadrupole multiplets has been reported 
for the semi-magic $N=82$ 
isotones \cite{bbd89,phk90,gvb90,gjb93}. 
These are states built on $3^{-}$ collective
octupole vibrational states.
However, most of the low-lying states of these isotones can be interpreted as 
pure excitations of valence protons outside a $^{132}$Sn closed core, and 
it offers a unique opportunity for studying the microscopic foundation of 
various nuclear models. 

Systematic studies of the $N=82$ isotones have been carried out by Andreozzi 
{\sl et al.} \cite{acgp90} who investigated the importance of pairing effects 
in these nuclei. Scholten {\sl et al.} \cite{sk83,ssbw87,swd89} have also 
studied the $N=82$ isotones and compared the generalized seniority 
scheme with the shell model. The most comprehensive shell model study, up to
now, was carried out by Wildenthal \cite{w90} for the $N=82$ isotones ranging
from $^{133}$Sb to $^{154}$Hf. The dimension of the 
problem was however reduced 
by both an occupation number and a seniority truncation. The 
effective two-body forces used in the above-mentioned 
shell model calculations 
are all rather schematic, phenomenological interactions.

One aim  of this work is to calculate an effective interaction based on
modern meson-exchange models for the nucleon-nucleon (NN) potential. 
The first step in the derivation of an effective interaction 
V$_{\rm eff}$ is to renormalize
the NN potential through the so-called $G$-matrix. The $G$-matrix is in turn
used in a perturbative many-body scheme discussed in Sect.\ 2  
to derive
an effective interaction for the $N=82$ isotones.

The second aim is to use this effective interaction 
in a full shell model (SM) calculation, 
within a model space or $P$-space consisting 
of the orbitals $2s_{1/2}$, $1d_{5/2}$, $1d_{3/2}$, $0g_{7/2}$ and $0h_{11/2}$ 
for the $Z=52 - 64$, $N=82$ isotones. This is the first time that a full shell
model calculation without any truncations has been performed for these
nuclei.

We have performed two similar and quite extended shell model calculations for 
the Sn isotopes, one having the doubly magic $^{100}$Sn as a closed core 
\cite{ehho95}, and  the other having the doubly magic $^{132}$Sn as a closed 
core \cite{ehhko95}. In the former, valence neutron particles have been added 
to the $^{100}$Sn core, and in the latter valence neutron particles have been 
subtracted from (or in other words, neutron holes have been added to)
the $^{132}$Sn core. In this work we present a further test of our 
method in the region of medium-heavy nuclei. In our previous works
on the Sn isotopes, we have carried out calculations based on the
neutron-neutron particle interaction and the neutron-neutron hole
interaction, respectively. In the present case an analogous test of the 
proton-proton particle interaction is provided by the $N=82$ isotones.

Our third aim is to compare the shell model results with results from a
quasiparticle random-phase approximation (QRPA) calculation.
We are performing the comparison with QRPA in order to test the perturbation
technique and to study to what extent the effective matrix elements give 
satisfactory results.
For the QRPA we take  as effective interaction the same $G$-matrix 
as the one used in the perturbative many-body scheme discussed above,
but the model space is enlarged by including the 
$1p0f$ shell as well.

The philosophy behind the perturbative approach 
is to include degrees of freedom 
not accounted for in the model space through 
various terms in perturbation theory. 
Therefore, since the QRPA calculation discussed here employs a larger 
single-particle 
space than the perturbative many-body scheme, 
the hope is that the two approaches 
can shed light on different many-body contributions and 
their influence on various 
spectroscopic observables.

This work is organized as follows:
In Sect. 2 we give a brief sketch on how to derive the effective interaction.
In the subsequent section some general spectroscopic features about the 
$N=82$ isotones are given together with a 
description of the shell model problem.
In Sect. 4, a short overview of the QRPA 
relevant for the $N=82$ isotones is given, 
whereas our results and discussions are presented in Sect. 5.
Concluding remarks are drawn in Sect. 6.

\section{Effective interaction for the $N=82$ isotones}

In nuclear structure calculations, we solve the quantum
many-body Schr\"{o}dinger equation for an $A$-nucleon system
\begin{equation}
     H\Psi_i(1,...,A)=E_i\Psi_i(1,...,A)
     \label{eq:full_a}
\end{equation}
in a restricted Hilbert space, referred to as the model space.
In Eq.\ (\ref{eq:full_a}), we have defined
$H=T+V$,
$T$ being the kinetic energy operator and $V$ the nucleon-nucleon (NN)
potential.
$E_i$ and $\Psi_i$ are the eigenvalues and eigenfunctions
for a state $i$ in the Hilbert space.
Introducing the auxiliary single-particle potential $U$, $H$ can
be rewritten as
\begin{equation}
    \begin{array}{ccc}H=H_{0}+H_1;&H_{0}=T+U;&H_1=V-U\end{array}.
\end{equation}
If $U$ is chosen such that $H_1$ becomes small, then $H_1$
can be treated as a perturbation.
The eigenfunctions of $H_{0}$
are then the unperturbed wave functions $\psi_i$.
These eigenfunctions can, in turn, be used to define a projection 
operator for the above-mentioned model space     
\begin{equation}
     P=\sum_{i=1}^{d}\left | \psi_i\right\rangle 
     \left\langle\psi_i\right | ,
\end{equation}
with $d$ being the size of the model space, and an excluded space
defined by the operator $Q$
\begin{equation}
     Q=\sum_{i=d+1}^{\infty}\left | \psi_i\right\rangle 
     \left\langle\psi_i\right | ,
\end{equation}
such that $PQ=0$.
The assumption then is that the most relevant components of the low-lying
nuclear states can be fairly well reproduced by configurations consisting
of few particles and holes occupying a limited number of orbitals $\psi_{i}$
selected on physical grounds.
These selected orbitals define the model space.

Eq.\ (\ref{eq:full_a})  can be rewritten as a secular equation
\begin{equation}
    PH_{\rm{eff}}P\Psi_i=P(H_{0}+V_{\rm{eff}})
    P\Psi_i=E_iP\Psi_i,
\end{equation}
where $H_{\rm{eff}}$  now is an effective hamiltonian acting solely
within the chosen model space. The definition of this effective interaction
is that it should act within the chosen model space and that the model-space
eigenvalue problem yields some of the
eigenvalues of the original hamiltonian. In general, however, 
these requirements do not determine the effective interaction
uniquely, as discussed in Refs.\ \cite{hko95,ko90,so95,sok94}. Several
many-body techniques exist for deriving the effective interaction
\cite{hko95,so95,sok94}. In this work we shall derive the effective
interaction using a time-dependent approach, starting from the time 
evolution operator $U(t,t')$ \cite{ko90}.
Our effective interaction is derived by the so-called folded-diagram 
expansion method of Kuo and co-workers \cite{ko90} (see below). The folded 
diagrams represent a set of diagrams which can be summed to infinite order 
through e.g.\ iterative methods. They arise when one removes the 
dependence on the exact energy of the perturbation expansion. 

Our scheme to obtain an effective interaction, appropriate for the $N=82$ 
isotones, starts with $A=132$ as the closed-shell core, and  can be divided 
into three steps. A  more detailed exposition can be found in Ref.\ 
\cite{hko95}.

First, one needs a free NN interaction $V$ which is
appropriate for nuclear physics at low and intermediate energies. At
present, a meson-exchange picture for the potential model seems to offer a
viable approach. Among such meson-exchange
models, one of the most successful is the one-boson-exchange model of the
Bonn group \cite{mac89}. As a starting point for our perturbative analysis, 
we shall use the parameters of the Bonn A potential defined in table A.1 of 
Ref.\ \cite{mac89}. For applications of this potential to 
nuclear structure, see e.g., Ref.\ \cite{mac89}.

In nuclear many-body calculations the first problem one is
confronted with is the fact that the repulsive core of the NN potential $V$
is unsuitable for perturbative approaches. This problem is overcome by 
the next
step in our many-body scheme, namely 
by introducing the reaction matrix $G$
\begin{equation}
    G=V+V\frac{\tilde{Q}}{\omega - H_0}G,
     \label{eq:g-matrix}
\end{equation}
where $\omega$ is the unperturbed energy of the interacting nucleons,
and $H_0$ is the unperturbed hamiltonian. 
The operator $\tilde{Q}$, commonly referred to
as the Pauli operator, is a projection operator which prevents the
interacting nucleons from scattering into states occupied by other nucleons.
In this work we solve the Bethe-Goldstone equation, using the so-called 
double-partitioning 
scheme \cite{hko95}, replacing $H_{0}$ in the denominatior 
of  Eq. (\ref{eq:g-matrix}) by $\tilde{Q} T \tilde{Q}$. 
To construct the Pauli operator which defines $G$, one has to take 
into account that neutrons and protons have different closed shell
cores, $N=82$ and $Z=50$, respectively.
This means that neutrons in the $2s1d0g_{7/2}0h_{11/2}$ shell
are holes, while protons in the $2s1d0g_{7/2}0h_{11/2}$ shell
are particles. For protons the  Pauli operator must be constructed so as to
prevent scattering into intermediate states with a single
proton in any of the
states defined by the orbitals from the $0s$ shell up to the
$0g_{9/2}$ orbital. For a two-particle state with protons only, one has 
also to avoid scattering 
into states with two protons in the $2s1d0g$ ($0g_{9/2}$ excluded)
and the $2p1f0h$ shells. For neutrons one must
prevent scattering into intermediate states with a single neutron
in the orbitals from the $0s$ shell up to the
$0h_{11/2}$ orbital. In addition, in case of a two-particle
state with neutrons only, one must prevent scattering into states
with two neutrons in the $0h_{9/2}0i_{13/2}1f2p$ shell
and the $3s2d1g0i_{11/2}0j_{15/2}$ shell. 
If we have a proton-neutron two-particle state we must in addition prevent
scattering into two-body states where a proton is in the 
the $2s1d0g$- ($0g_{9/2}$ excluded)
and the $2p1f0h$ shells and a neutron is in the 
$0h_{9/2}0i_{13/2}1f2p$ shell
and the $3s2d1g0i_{11/2}0j_{15/2}$ shell.
The single-particle wave functions were chosen to be harmonic oscillator 
eigenstates with the oscillator energy 
$\hbar\Omega = 45A^{-1/3} - 25A^{-2/3}=7.87 $ MeV,  for $A=132$.
This $G$-matrix will also be used as the effective
interaction in the QRPA calculations discussed
in the next section. 

The last step consists in defining a
two-body interaction in terms of the $G$-matrix.
The first step here is to define the so-called $\hat{Q}$-box given by
\begin{equation}
   P\hat{Q}P=PH_1P+
   P\left(H_1 \frac{Q}{\omega-H_{0}}H_1\\ +H_1
   \frac{Q}{\omega-H_{0}}H_1 \frac{Q}{\omega-H_{0}}H_1 +\dots\right)P,
   \label{eq:qbox}
\end{equation}
where we will replace $H_1$ with $G-U$ ($G$ replaces the free NN interaction
$V$).
The $\hat{Q}$-box is made up of non-folded diagrams which are irreducible
and valence linked.
A diagram is said to be irreducible if between each pair
of vertices there is at least one hole state or a particle state outside
the model space. In a valence-linked diagram the interactions are linked
(via fermion lines) to at least one valence line. Note that a valence-linked
diagram can be either connected (consisting of a single piece) or
disconnected. In the final expansion, including folded diagrams as well, the
disconnected diagrams are found to cancel out \cite{ko90}.
This corresponds to the cancellation of unlinked diagrams
of the Goldstone expansion \cite{ko90}.
The projection operator $Q$ used in the definition 
of the effective interaction need not be the same as $\tilde{Q}$
 used in 
the calculation of the $G$-matrix. This is the case in our calculation
since we are using the double-partitioning scheme for calculating the
$G$-matrix. Such an approach leads to the inclusion of additional
ladder diagrams in the definition of the $\hat{Q}$-box, as discussed in
Ref.\ \cite{hko95}.
We obtain the effective interaction,
$H_{\rm{eff}}=H_0+V_{\rm{eff}}$, in terms of the $\hat{Q}$-box as
\cite{hko95,ko90}
\begin{equation}
    V_{\rm{eff}}^{(n)}=\hat{Q}+{\displaystyle\sum_{m=1}^{\infty}}
    \frac{1}{m!}\frac{d^m\hat{Q}}{d\omega^m}\left\{
    V_{\rm{eff}}^{(n-1)}\right\}^m.
    \label{eq:fd}
\end{equation}
Observe also that the
effective interaction $V_{\rm{eff}}^{(n)}$
is evaluated at a given model space energy
$\omega$, as is the case of the $G$-matrix. Here we choose
$\omega =-20$ MeV, although the dependence
of the final resulting spectra of the choice
of starting energy is rather weak. Since the higher-order derivatives of the 
$\hat{Q}$-box are rather small, the series can be truncated 
at $m\sim 6-10$, and similarly some $6-10$ iterations $n$ are 
needed for convergence of the effective interaction. 
The $\hat{Q}$-box in this work is defined to be the sum
of all non-folded diagrams through third order in the $G$-matrix, as
discussed in Ref.\ \cite{hko95}.

\section{Shell model calculation}

For the shell model calculation we define the model space to consist of the 
spherical single-particle orbitals in the $N=4$ oscillator shell 
($1d_{5/2}$, $0g_{7/2}$, $1d_{3/2}$, $2s_{1/2}$) plus the intruder $0h_{11/2}$ 
orbital from the $N=5$ oscillator shell. Hereafter this model space is 
referred to as the $sdg$-shell. This means that our $P$-space consists of the
proton orbitals outside the 
$^{132}$Sn core, ranging from the closed $Z=50$, $N=82$ core to the closed 
$Z=N=82$ core. 
 
\begin{figure}[htbp]
\setlength{\unitlength}{1.4cm}

\begin{center}
\begin{picture}(2,5)(0,-1)
\newcommand{\lc}[1]{\put(0,#1){\line(1,0){1}}}
\newcommand{\ls}[2]{\put(2,#1){\makebox(0,0){{\scriptsize $#2$}}}}
\newcommand{\lsr}[2]{\put(2,#1){\makebox(0,0){{\scriptsize $#2$}}}}
\put(-.25,3.4){\makebox(0,0){\large MeV}}
\thicklines
\put(-.75,-.5){\line(0,1){4}}
\multiput(-.75,.0)(0,1){4}{\line(1,0){.1}}
\multiput(-.75,.5)(0,1){3}{\line(1,0){.05}}
\put(-1.,3){\makebox(0,0){3}}
\put(-1.,2){\makebox(0,0){2}}
\put(-1.,1){\makebox(0,0){1}}
\put(-1.,0){\makebox(0,0){0}}
\lc{0.000}   \ls{0.000000}{7/2+ \;\;0.000}
\lc{0.962}   \ls{0.962}{5/2+ \;\;0.962}
\lc{2.708}   \ls{2.608}{3/2+ \;\;2.708}
\lc{2.792}   \ls{2.792}{11/2- \;\;2.792}
\lc{2.990}   \ls{2.990}{1/2+ \;\;2.990}
\end{picture}
\end{center}

\caption{Adopted single-particle energies for the orbitals $2s_{1/2}$,
$1d_{3/2}$, $1d_{5/2}$, $0g_{7/2}$ and $h_{11/2}$ in the shell model 
calculation.}
\label{fig:sp-energies}
\end{figure}
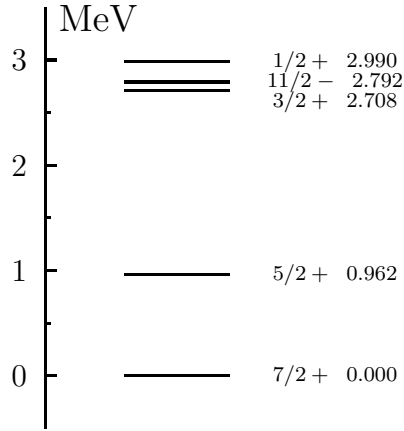

At present it does not seem possible to 
calculate the $P$-space single-particle
energies along the same lines as the effective two-body interaction. 
The theoretical framework is available, but the results are not accurate
enough for our purpose. However, the single-particle energies can be
extracted from the experimental $^{133}$Sb spectrum \cite{stone}, except 
for the $2s_{1/2}$ single-particle state which has not yet been measured. 
For the $2s_{1/2}$ single-particle energy we have taken the value used by 
Sagawa {\sl et al.} \cite{ssbw87}. The adopted single-particle energies are 
as displayed in Fig.\ \ref{fig:sp-energies}.

In the shell model calculation, all degrees of freedom within the defined 
$P$-space are included. Thus, the dimension of the problem grows rapidly with
increasing number of valence particles as shown in Table \ref{tab:dimension}.
Our basic approach to solving the many-body eigenvalue problem is the Lanczos
algorithm, a method which was first applied to nuclear physics problems by 
Whitehead 
{\sl et al.} \cite{whit77}. This is an iterative method, where the $10-20$
low-lying eigenstates of interest are 
obtained after a rather limited number of 
iterations. The shell model algorithm used is reviewed in more detail in 
Ref.\ \cite{ehho95}.

\begin{table}[htbp]
\begin{center}
\caption{Number of basis states for the shell model calculation of the $N=82$
isotones, with $1d_{5/2}$, $0g_{7/2}$, $1d_{3/2}$, $2s_{1/2}$ and $0h_{11/2}$
single particle orbitals.}
\begin{tabular}{lr|cr|lr}
\\\hline
System & Dimension & System & Dimension & System & Dimension \\
\hline
$^{134}$Te & 36        & $^{139}$La & 108 297   & $^{144}$Sm & 6 210 638 \\
$^{135}$I  & 245       & $^{140}$Ce & 323 682   & $^{145}$Eu & 9 397 335 \\ 
$^{136}$Xe & 1 504     & $^{141}$Pr & 828 422   & $^{146}$Gd & 12 655 280 \\
$^{137}$Cs & 7 451     & $^{142}$Nd & 1 853 256 & $^{147}$Tb & 15 064 787 \\
$^{138}$Ba & 31 124    & $^{143}$Pm & 3 609 550 & $^{148}$Dy & 16 010 204 \\
\hline
\end{tabular}
\label{tab:dimension}
\end{center}
\end{table}

\begin{figure}[htbp]
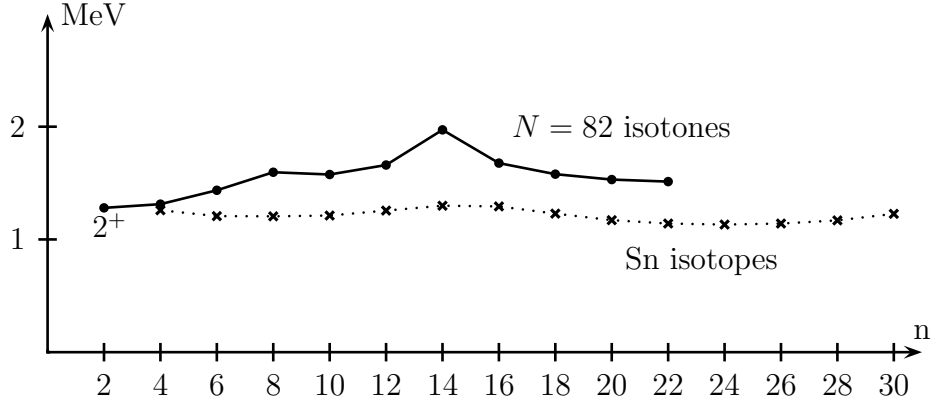

 \setlength{\unitlength}{1cm}
 \begin{center}
 \setlength{\unitlength}{1cm}
 \thicklines
 %
%

 \Cartesian(0.75cm,1.5cm)
 \pspicture(-1,-1)(16,3.5)
\psaxes[Ox=0,Dx=2,dx=1,showorigin=false,linewidth=1pt]{->}(0,0)(15.5,3.0)
 \uput[0](0.0,3.0){MeV}
 \uput[0](8.0,2.0){$N=82$ isotones}
 \uput[0](10.0,0.8){Sn isotopes}
 \uput[90](15.5,0.0){n}
 %
 %
 %
\psline[showpoints=true,linestyle=dotted,dotstyle=+,dotangle=45,
dotscale=1.2,linewidth=1pt]
(2,1.259)(3,1.207)(4,1.205)(5,1.212)(6,1.256)(7,1.299)(8,1.293)
(9,1.229)(10,1.171)(11,1.141)(12,1.132)(13,1.141)(14,1.169)(15,1.227)
\psline[showpoints=true,linestyle=solid,dotstyle=*,dotscale=1.2,linewidth=1pt]
 (1,1.280)(2,1.313)(3,1.436)(4,1.596)(5,1.576)(6,1.660)(7,1.972)(8,1.677)
 (9,1.579)(10,1.531)(11,1.513)
\uput[90](1.1,0.9){$2^{+}$}
\endpspicture
%
 %
 \end{center}
 \caption{\label{fig:2+-states}Experimental excitation energies for the lowest
 $2^{+}$ state in the Sn isotopes compared with the $N=82$ isotones, where $n$
 is the number of valence particles relative to $^{100}$Sn 
and $^{132}$Sn respectively. }
 \end{figure}

For the low energy region of the Sn isotopes where the main degrees of 
freedom are valence neutrons filling up the $sdg$-shell, a characteristic 
feature of the even-even nuclei is the remarkably constant spacing between 
the $0^{+}$ ground state and the first excited $2^{+}$ state. The $N=82$ 
isotones have a different closed core, but what still should be important is
the filling of the $sdg$-shell, now with protons. In Fig.\
\ref{fig:2+-states} we have compared the empirical $0^{+}_{1} - 2^{+}_{1}$ 
spacing for the Sn isotopes and the $N=82$ isotones. 
As can be seen, the 
$N=82$ isotones do not have the same type of constant $0^{+}_{1} -2^{+}_{1}$ 
spacing as is observed for the Sn isotopes. The spacing increases slightly 
with increasing proton number until the middle of the shell at $Z=64$ where 
there is a sudden increase in the spacing yielding a gap, close to $2$ MeV in 
magnitude.

Small but important differences in 
the single-particle spectra are found between 
the Sn isotopes and the $N=82$ isotones. The strong pairing effect which is 
seen for tin, is due to the nearly degenerate $1d_{5/2}$ and $0g_{7/2}$ 
single-particle orbitals. These orbitals are predicted, from a shell model 
extrapolation by Grawe {\em et al.} \cite{gsm95}, to be approximately $0..20$ 
MeV apart. From $^{133}$Sb we know that the $1d_{5/2}$ and $0g_{7/2}$ 
orbitals are interchanged relative to $^{101}$Sn, and separated by $0.96$ MeV.
The larger separation of the two orbitals may explain the less stable 
$0^{+}_{1} - 2^{+}_{1}$ spacing in the $N=82$ isotones.

Much attention has been devoted to $^{146}$Gd and to the question 
to what extent $^{146}$Gd can be considered as a stable closed subshell 
nucleus. Similarities between $^{146}$Gd and $^{208}$Pb, like the $3^{-}$
first excited state and the large gap between the $0^{+}$ ground
state and the first excited $2^{+}$ state, led to intensive studies
\cite{kbd79,obd78,nsp81} and speculations about the size of the 
single-particle energy gap at $Z=64$. However, it now seems clear that 
$^{146}$Gd has doubly magic properties, but the subshell closure does not 
appear as pronounced as for $^{208}$Pb \cite{wsg89}. 

Above we have described some general, qualitative properties of the $N=82$
isotones, and it is essential that our microscopic shell model calculation is 
able to reproduce such phenomena.

\section{The $N=82$ isotones in the framework of the QRPA}

The quasiparticle random-phase approximation (QRPA) is here presented as an
alternative method to the shell model.
The basic ingredients of the QRPA approach used in this work are described 
in Ref.\ \cite{suh94}. 
We start by solving the BCS equations for both protons 
and neutrons in a larger space of single-particle orbitals than included in 
the shell model calculation. Then a QRPA energy 
matrix is calculated and solved for each spin separately.
The nuclear states are obtained as linear combinations of proton-proton and 
neutron-neutron two-quasiparticle excitations.
A bare $G$-matrix is used for the interaction and the idea is that 
quasiparticle excitations replace 
the effect of the perturbation calculation in 
the shell model case. Thus, instead of treating core excitations in 
perturbation theory producing an effective interaction between valence protons,
the quasiparticle excitations are explicitly included in the nuclear wave 
functions. New states may be obtained not present in the shell model approach 
if a large part of the corresponding wave functions are core excitations. 
Such configurations are not treated properly in perturbation theory.
In the present case the proton single-particle
basis is taken to consist of the $1p0f$ and $2s1d0g$ oscillator major shells
complemented with the $0h_{11/2}$ intruder state from the major shell 
above. This leaves $Z=20$ as an inert proton core.
For neutrons we have chosen the 
valence space to consist of the $2s1d0g_{7/2}$ and $2p1f0h$ major shells 
leaving $N=50$ as the neutron core. 

In the QRPA calculation,
the two-body interaction used to construct the energy matrix,
is obtained from the Bonn one-boson-exchange potential giving the same 
nuclear $G$-matrix discussed in Sect.\ 2. 
In this calculation the bare $G$-matrix has been applied and 
no attempts are made to create a 
model-space adapted effective interaction
because the single-particle space for the QRPA is relatively large.
In order to have as equal as possible starting points in the QRPA and
the shell model approaches, we use the proton single-particle
energies as shown in Fig.\ \ref{fig:sp-energies}. For protons in 
the $1p0f0g_{9/2}$ shell and neutrons in the $2s1d0g_{7/2}0h_{11/2}$
and $1p0f0g_{9/2}$ shells we employ Woods-Saxon single-particle
energies calculated using the parametrization of Bohr and Mottelson
\cite{boh63}. 

The QRPA calculation is slightly different from the more traditional
ones as discussed
in Ref.\ \cite{suh94}.
It has been 
pointed out in Ref.\ \cite{acgp90} that pairing effects are strong 
in the $N=82$ isotones and this means 
that it is essential to exploit the 
available experimental data on pairing gaps and single-quasiparticle energies 
if one wants to 
obtain an improved QRPA description of the low-energy properties of the 
nuclei under study. 
Typically, certain matrix elements of the $G$-matrix may be scaled
in order to reproduce the systematics of pairing in certain 
isotopes and isotones, as done in Refs.\ \cite{suh94,suh95,ber95}.
However, in this work we have not
performed such a refitting of the nuclear $G$-matrix.
If one adjusts the $G$-matrix by some scaling constants in order
to have a better reproduction of the spectra in the QRPA one introduces
many-body effects whose origin are difficult to retrace. Although
the QRPA may not give the best results for the spectroscopy, 
we feel that an equal starting point, same $G$-matrix and single-particle
energies, may offer a better possibility
for studying differences between the QRPA and the shell model.

 From the present discussion and the one in Sect. 2 it is obvious that
the QRPA approach exhibits two important differences compared
to the shell model approach.
First, the single-particle basis
for protons is larger for the QRPA allowing for proton
core excitations across the $Z=50$ shell gap. Second, also the neutrons are
active  yielding neutron core excitations across the $N=82$ shell gap. This
might become important for the description of some low-energy collective
excitations of the even $N=82$ isotones. In the shell model approach
these degrees of freedom are supposed to be accounted for by terms
included in the perturbative expansion of the effective interaction.
Substantial differences in the two approaches may therefore
reveal whether such low-energy collective excitations are accounted
for in the shell model approach where the calculations are
done within a smaller single-particle space, but with a complete
set of many-body basis states within the chosen model space.

\section{Numerical results and discussion}

\subsection{Energy levels}

The calculated energy eigenstates of the QRPA and the shell model (SM) along
with the experimental energy levels are presented in Tables\ \ref{tab:even-1}, 
\ref{tab:even-2} and \ref{tab:even-3}. If nothing else is specified data 
are taken from the data base of the National Nuclear Data Center, Brookhaven
National Laboratory, Upton, N.Y., USA.
A large number of E2 transitions are experimentally
known for the $N = 82$ isotones. Data for some selected transitions are 
presented in Table \ref{tab:even-e2-1} and compared with our theoretical 
predictions. Some E3 trasitions are also presented in the same table.

In the SM calculation of E2 transitions an effective proton charge 
$e_{\rm p}^{\rm eff}({\rm SM}) = 1.4e$ is used in 
agreement with the discussion in \cite{boh63}. 
Our model space which includes the 
$2s1d0g_{7/2}0h_{11/2}$ single-particle orbitals
would require an 
effective E2 operator which may be calculated along the same lines
as the effective interaction discussed 
in Sect.~2. However, at present we
limit ourselves to a constant effective E2 charge. For the QRPA 
a larger single-particle space is explicitly included in the  calculation
and an  effective E2 charge $e_{\rm p}^{\rm eff}({\rm QRPA}) = 1.0e$ 
is appropriate.

There are two important questions related to the present calculation.
First, can a medium dependent
effective interaction which  is calculated starting
from the free nucleon-nucleon (NN)
potential
be used in medium-heavy nuclei? Secondly, which states in the N = 82
isotones are well described by the chosen model space?
As discussed in Sect.\ 2, our approximations lie in the choice of
model space and selection of many-body diagrams in
the definition of the effective interaction.
The only parameters which enter our theory are those 
which define the NN potential. 
Thus, in case of disagreements with observation it is important
not to modify the derived effective interaction
in order to get an improved reproduction of the data. 
Such disagreements may point to degrees of freedom not accounted for 
in our many-body scheme.

In view of these restrictions our results are rather good.
For both models the deviation of our calculated 
energy levels from the experimental ones is generally within $0.1 - 0.3$ 
MeV. Up to five 2$^+$ states, three 4$^+$ states and two 6$^+$ states
are well reproduced throughout the sequence of isotones
indicating that the degrees of freedom represented by the chosen proton 
model space are the relevant ones.   
A decomposition of the QRPA $2^{+}_{1}$ wave functions shows
that these states are constructed mainly of two-quasiparticle proton
excitations within the $sdg$-shell, a picture which is consistent with the SM. 
There is a very  small contribution of neutron core excitations in the 
wave functions, a contribution 
which increases slightly towards the middle of the shell.
Furthermore,  
the high-spin states 8$^+$ and 10$^+$ are well reproduced
except for  $^{144}$Sm where the theoretical states 
are too high in energy (0.58~MeV for 8$^{+}$ and 0.97~MeV for 10$^{+}$).
This may indicate that other degrees of freedom are important for these 
high spin states. 

With one exception the known E2 transitions are well reproduced.
A  typical feature of the even $N=82$ isotones is enhanced E2 
transitions between the first excited $2^{+}$ state and the $0^{+}$ 
ground state with strengths around $10$ W.u. 
These transitions strengths are well reproduced  by the proton degrees of freedom 
within the SM valence space with some additional polarization charge which can be described 
in perturbation theory.  
This effect is  also seen in the structure of the $2^{+}_{1}$ state in the QRPA framework, 
where a large number of 
two-quasiparticle components are present and contribute coherently
to the E2 transition rate.
We obtain $4^{+}_{1} \rightarrow 2^{+}_{1}$ and 
$2^{+}_{1} \rightarrow 0^{+}_{1}$ E2 transitions in good agreement with 
experiment. The $6^{+}_{1} \rightarrow 4^{+}_{1}$ transition in $^{134}$Te is 
also reasonably well reproduced. For the other isotones this transition is 
weak, and we have more difficulties in reproducing these data.
In spite of the rather large deviation in energy between the calculated 
$8^{+}_{1}$ and $10^{+}_{1}$ states and the corresponding experimental states, 
the E2 $10^{+}_{1} \rightarrow 8^{+}_{1}$ values are in good agreement with 
data.

\begin{figure}[htbp]
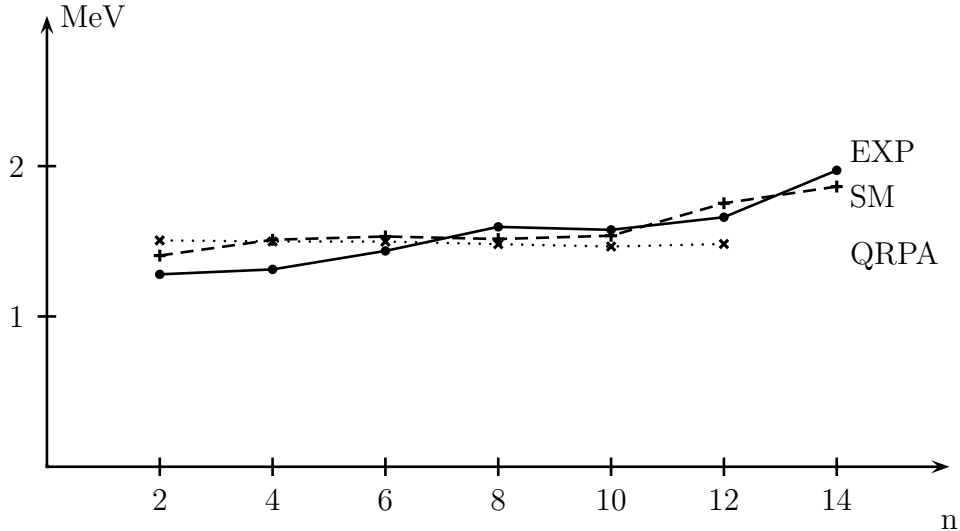

 \setlength{\unitlength}{1cm}
 \begin{center}
 \setlength{\unitlength}{1cm}
 \thicklines
 \Cartesian(1.5cm,2cm)
 \pspicture(-1,-.5)(8,3)
\psaxes[Ox=0,Dx=2,dx=1,showorigin=false,linewidth=1pt]{->}(0,0)(8,3.0)
 \uput[0](0.0,3.0){MeV}
 \uput[0](7,2.1){EXP}
 \uput[0](7,1.8){SM}
 \uput[0](7,1.4){QRPA}

 \uput[90](8,-.5){n}
 %
 %
 %
 %
\psline[showpoints=true,linestyle=solid,dotstyle=*,dotscale=1.2,linewidth=1pt]
 (1,1.280)(2,1.313)(3,1.436)(4,1.596)(5,1.576)(6,1.660)(7,1.972)
%
\psline[showpoints=true,linestyle=dashed,dotstyle=+,dotscale=1.2,linewidth=1pt]
 (1,1.405)(2,1.510)(3,1.532)(4,1.515)(5,1.537)(6,1.753)(7,1.864)
%
\psline[showpoints=true,linestyle=dotted,dotstyle=+,dotangle=45,dotscale=1.2,linewidth=1pt]
 (1,1.506)(2,1.499)(3,1.498)(4,1.481)(5,1.465)(6,1.482)
\endpspicture
%
 %
 \end{center}
 \caption{\label{fig:2states}The $2^{+}_{1}$ state in the $N=82$ isotones, 
 where $n$ is the number of valence particles relative to $^{132}$Sn. }
 \end{figure}

An SM calculation for $^{146}$Gd is difficult due to the large number 
of basis states, see  Table~\ref{tab:dimension}. At present we have 
only calculated the ground state and the first excited 2$^{+}$ state.
The $0^{+}_{1}-2^{+}_{1}$ spacing is found to be $1.864$ MeV and compares 
well to the experimental  value of $1.972$~MeV.
However, the experimental increase in energy for the 2$^{+}$ state from $^{144}$Sm to $^{146}$Gd
indicates shell closure of the $d_{5/2}g_{7/2}$ single-particle orbits.
The SM calculation reproduces this feature but not as sharp as found experimentally. 
The  $0^{+}_{1}-2^{+}_{1}$ spacing throughout the sequence of isotones is shown
in Fig.~\ref{fig:2states}.
The SM reproduces the weak increase resonably well but with 
some small deviations at the endpoints. The QRPA predicts 
a constant spacing due to the use of the BCS approximation.

\begin{table}[htbp]
\begin{center}
\caption{Low-lying states for $^{134}$Te and $^{136}$Xe. Experimental 
angular
momentum values in parentheses are tentative. Energies are given in MeV.}
\begin{tabular}{ccccc|ccccc}
\hline
\multicolumn{5}{c|}{$^{134}$Te} & \multicolumn{5}{c}{$^{136}$Xe} \\
$J^{\pi}$ & QRPA & SM & Exp & & $J^{\pi}$ &  
QRPA & SM & Exp & \\ 
\hline
$0^{+}_{2}$ & 1.936 & 2.715 &  & & 
$0^{+}_{2}$ & 1.414 & 2.174 & 2.582 &$0^{+}$ \\
$0^{+}_{3}$ & 2.686 & 6.316 & & & 
$0^{+}_{3}$ & 2.387 & 3.049 &  & $0^{+}$ \\
$0^{+}_{4}$ & 3.876 & 7.141 &  & & 
$0^{+}_{4}$ & 4.124 & 3.762 & 4.320& \\
$2^{+}_{1}$ & 1.506 & 1.405 & 1.279 & 
$2^{+}$ &$2^{+}_{1}$ & 1.499 & 1.510 & 1.313 & $2^{+}$ \\
$2^{+}_{2}$ & 2.619 & 2.646 & 2.464 & $(2^{+})$ &
$2^{+}_{2}$ & 2.415 & 2.382 & 2.290 & $2^{+}$ \\
$2^{+}_{3}$ & 2.928 & 3.258 & 2.934 & $(2^{+})$ &
$2^{+}_{3}$ & 2.465 & 2.674 & 2.415 & $2^{+}$ \\
$2^{+}_{4}$ & 2.992 & 4.063 & & &
$2^{+}_{4}$ & 2.720 & 2.706 & 2.634 & $2^{+}$ \\
$2^{+}_{5}$ & 3.493 & 5.196 &  & &
$2^{+}_{5}$ & 3.300 & 2.944 & 2.849 & $2^{(+)}$ \\
$3^{+}_{1}$ & 2.006 & 2.793 & & &
$3^{+}_{1}$ & 2.137 & 2.505 & 2.126 & $3^{+},4^{+}$ \\ 
$3^{-}_{1}$ & 3.374 &       & & &
$3^{-}_{1}$ & 3.095 &       & 3.275 & $3^{-}$ \\ 
$3^{-}_{2}$ & 4.425 & 4.313 & & &
$3^{-}_{2}$ & 4.181 & 3.992 & & \\
$4^{+}_{1}$ & 1.849 & 1.717 & 1.570 & $4^{+}$ & 
$4^{+}_{1}$ & 1.956 & 1.895 & 1.694 & $4^{+}$ \\
$4^{+}_{2}$ & 2.588 & 2.701 & & & 
$4^{+}_{2}$ & 2.463 & 2.400 & 2.465 & $(4^{+})$ \\
$4^{+}_{3}$ & 3.190 & 3.548 & & &
$4^{+}_{3}$ & 2.744 & 2.575 & 2.560 & $4^{+}$ \\
$5^{+}_{1}$ & 2.006 & 2.826 & 2.727 & $(5^{+})$ &
$5^{+}_{1}$ & 2.137 & 2.448 & 2.444 & $5$ \\ 
$6^{+}_{1}$ & 1.996 & 1.893 & 1.691 & $6^{+}$ &
$6^{+}_{1}$ & 2.110 & 2.095 & 1.892 & $6^{+}$ \\  
$6^{+}_{2}$ & 2.348 & 2.394 & 2.396 & $(6^{+})$ &
$6^{+}_{2}$ & 2.266 & 2.166 & 2.262 & $6^{+}$ \\
$8^{+}_{1}$ & 3.424 & 7.362 & 4.557 & $8^{+}$ &
$8^{+}_{1}$ & 2.914 & 3.327 &  & \\
$10^{+}_{1}$ &3.524 & 7.538 & 5.622 & $10^{+}$ &
$10^{+}_{1}$ &3.014 & 3.671 & & \\
\hline
\end{tabular}
\label{tab:even-1}
\end{center}
\end{table}

\begin{table}[htbp]
\begin{center}
\caption{Low-lying states for $^{138}$Ba and $^{140}$Ce. Experimental 
angular
momentum values in parentheses are tentative. Energies are given in MeV.}
\begin{tabular}{ccccc|ccccc}
\hline
\multicolumn{5}{c|}{$^{138}$Ba} & \multicolumn{5}{c}{$^{140}$Ce} \\
$J^{\pi}$ & QRPA & SM & Exp & & $J^{\pi}$ &  
QRPA & SM & Exp & \\ 
\hline
$0^{+}_{2}$ & 1.246 & 2.105 & 2.340 & $0^{+}$ &
$0^{+}_{2}$ & 2.209 & 2.234 & 1.903 & $0^{+}$ \\
$0^{+}_{3}$ & 2.210 & 2.978 & 3.612 & $(0^{+})$ &
$0^{+}_{3}$ & 2.246 & 3.054 & 3.017 & $0^{+}$ \\
$0^{+}_{4}$ & 4.126 & 3.838 & 3.484 &         &
$0^{+}_{4}$ & 4.146 & 3.589 & 3.226 & $0^{+}$ \\
$2^{+}_{1}$ & 1.498 & 1.532 & 1.436 & $2^{+}$ &
$2^{+}_{1}$ & 1.481 & 1.515 & 1.596 & $2^{+}$ \\
$2^{+}_{2}$ & 2.257 & 2.381 & 2.190 & $(1,2,3)$ &
$2^{+}_{2}$ & 2.412 & 2.447 & 2.348 & $2^{+}$ \\
$2^{+}_{3}$ & 2.383 & 2.564 & 2.218 & $2^{+}$ &
$2^{+}_{3}$ & 2.617 & 2.630 & 2.521 & $2^{+}$ \\
$2^{+}_{4}$ & 2.587 & 2.862 & 2.583 & $(1^{+},2^{+})$ &
$2^{+}_{4}$ & 3.086 & 4.724 & 2.900 & $2^{+}$ \\
$2^{+}_{5}$ & 3.164 & 3.003 & 2.640 & $2^{+}$ &
$2^{+}_{5}$ & 3.255 & 6.042 & 3.001 & $2^{+}$ \\
$3^{+}_{1}$ & 2.347 & 2.479 & 2.446 & $3^{+}$ &
$3^{+}_{1}$ & 2.510 & 2.557 & 2.412 & $3^{+}$ \\
$3^{-}_{1}$ & 2.737 &       & 2.881 & $3^{-}$ &
$3^{-}_{1}$ & 2.320 &       & 2.464 & $3^{-}$ \\
$3^{-}_{2}$ & 3.919 & 3.699 & 3.647 & $(3)^{-}$ &
$3^{-}_{2}$ & 3.704 &       & 3.040 & $3^{-}$ \\
$3^{-}_{3}$ & 4.789 & 4.118 & 3.923 & $(3)^{-}$ &
$3^{-}_{3}$ & 4.865 & 3.484 & 3.473 & $3^{-}$ \\
$4^{+}_{1}$ & 2.080 & 2.035 & 1.898 & $4^{+}$ &
$4^{+}_{1}$ & 2.137 & 2.134 & 2.083 & $4^{+}$ \\
$4^{+}_{2}$ & 2.442& 2.422 & 2.308 & $4^{+}$ &
$4^{+}_{2}$ & 2.506 & 2.506 & 2.481 & $4^{+}$ \\
$4^{+}_{3}$ & 2.589 & 2.512 & 2.583 & $4^{+}$ &
$4^{+}_{3}$ & 2.608 & 2.542 & 2.516 & $3^{+},4^{+}$ \\
$4^{+}_{4}$ & 3.606 & 2.814 & 2.779 & $4^{+}$ &
$4^{+}_{4}$ & 3.292 & 4.701 & 3.331 & $4^{+}$ \\
$5^{+}_{1}$ & 2.347 & 2.421 & 2.415 & $5^{+}$ &
$5^{+}_{1}$ & 2.510 & 2.516 & 2.350 & $5^{+}$ \\
$6^{+}_{1}$ & 2.155 & 2.178 & 2.091 & $6^{+}$ &
$6^{+}_{1}$ & 2.211 & 2.238 & 2.108 & $6^{+}$ \\
$6^{+}_{2}$ & 2.386 & 2.425 & 2.203 & $6^{+}$ &
$6^{+}_{2}$ & 2.667 & 2.588 & 2.629 & $6^{+}$ \\
$8^{+}_{1}$ & 2.759 & 3.424 & 3.184 & $8^{+}$ &
$8^{+}_{1}$ & 3.761 & 3.476 & 3.513 & $8^{+}$ \\
$10^{+}_{1}$ & 2.859 & 3.904 & 3.622& $10^{+}$ &
$10^{+}_{1}$ & 3.861 & 4.065 & 3.715 & $10^{+}$ \\
\hline
\end{tabular}
\label{tab:even-2}
\end{center}
\end{table}

\begin{table}[htbp]
\begin{center}
\caption{Low-lying states for $^{142}$Nd and $^{144}$Sm. Experimental 
angular
momentum values in parentheses are tentative. Energies are given in MeV.}
\begin{minipage}[t]{5in}
\begin{tabular}{ccccc|ccccc}
\hline
\multicolumn{5}{c|}{$^{142}$Nd} & \multicolumn{5}{c}{$^{144}$Sm} \\
$J^{\pi}$ & QRPA & SM & Exp & & $J^{\pi}$ &  
QRPA & SM & Exp & \\ 
\hline
$0^{+}_{2}$ & 2.402 & 2.377 & 2.217 & $0^{+}$ &
$0^{+}_{2}$ & 1.947 & 2.397 & 2.478 & $0^{+}$ \\
$0^{+}_{3}$ & 3.074 & 2.674 & 2.978 & $0^{+}$ &
$0^{+}_{3}$ & 2.675 &       & 2.827 & $0^{+}$ \\
$0^{+}_{4}$ & 3.457 & 3.653 & 3.583 & $(0^{+})$ &
$0^{+}_{4}$ & 2.896 &       & 3.142 & $0^{+}$ \\
$2^{+}_{1}$ & 1.465 & 1.537 & 1.576 & $2^{+}$ &
$2^{+}_{1}$ & 1.482 & 1.753 & 1.660 & $2^{+}$ \\
$2^{+}_{2}$ & 2.516 & 2.510 & 2.385 & $2^{+}$ &
$2^{+}_{2}$ & 2.577 & 2.536 & 2.423 & $2^{+}$ \\
$2^{+}_{3}$ & 2.841 & 2.676 & 2.846 & $2^{+}$ &
$2^{+}_{3}$ & 2.610 & 2.639 & 2.661 & 
$(2^{+})$\footnote[1]{Ref.\ \cite{gjb93}} \\
$2^{+}_{4}$ & 3.070 &       & 3.046 & $(2)^{+}$ &
$2^{+}_{4}$ & 3.123 & 2.793 & 2.799 & $2^{+}$ \\
$2^{+}_{5}$ & 3.375 &       & 3.128 & $(1,2^{+})$ &
$2^{+}_{5}$ & 3.249 &       & 3.318 & $2^{+}$ \\
$3^{+}_{1}$ & 2.428 & 2.644 & 2.548 & $3^{+}$ &
$3^{+}_{1}$ & 2.528 &       & 2.687 & $3^{(+)\,{\rm a})}$ \\
$3^{-}_{1}$ & 1.882 &       & 2.085 & $3^{-}$ &
$3^{-}_{1}$ & 1.477 &       & 1.810 & $3^{-}$ \\
$3^{-}_{2}$ & 3.547 & 3.325 & 3.366 & $(3)^{-}$ &
$3^{-}_{2}$ & 3.446 & 3.196 & 3.228 & $3^{-}$ \\
$4^{+}_{1}$ & 2.140 & 2.235 & 2.101 & $4^{+}$ &
$4^{+}_{1}$ & 2.196 & 2.390 & 2.191 & $4^{+}$ \\
$4^{+}_{2}$ & 2.593 & 2.642 & 2.438 & $4^{+}$ &
$4^{+}_{2}$ & 2.771 & 2.737 & 2.588 & $4^{+}$ \\
$4^{+}_{3}$ & 2.932 & 2.667 & 2.738 & $4$ &
$4^{+}_{3}$ & 2.787 & 2.830 & 2.884 & $4^{+}$ \\
$5^{+}_{1}$ & 2.428 & 2.626 & 2.514 & $5^{+}$ &
$5^{+}_{1}$ & 2.528 &       & 2.704 & $(5^{+})^{\,{\rm a})}$ \\
$6^{+}_{1}$ & 2.364 & 2.316 & 2.210 & $6^{+}$ &
$6^{+}_{1}$ & 2.656 & 2.363 & 2.323 & $6^{+}$ \\
$6^{+}_{2}$ & 3.062 &       & 2.887 & $6^{+}$ &
$6^{+}_{2}$ & 2.875 &       & 2.729 & $(6^{+})$ \\
$8^{+}_{1}$ & 4.349 & 3.819 & 3.454 & $8^{+}$ &
$8^{+}_{1}$ & 3.636 & 4.193 & 3.651 & $8$ \\
$10^{+}_{1}$ & 4.451 & 4.080 & 3.926 & $10^{+}$ &
$10^{+}_{1}$ & 3.735 & 5.195 & 4.221 & $10$ \\
\hline
\end{tabular}
\end{minipage}
\label{tab:even-3}
\end{center}
\end{table}

\begin{table}[htbp]
\begin{center}
\caption{E2 and E3 transitions for $^{134}$Te - $^{144}$Sm. For theory 
$e_{\rm p}^{\rm eff}({\rm SM}) = 1.4e$ and 
$e_{\rm p}^{\rm eff}({\rm QRPA}) = 1.0e$ for E2 transitions, while
$e_{\rm p}^{\rm eff}({\rm SM}) = 1.0e$ and 
$e_{\rm p}^{\rm eff}({\rm QRPA}) = 1.0e$ for E3 transitions. 
All entries are given in 
Weisskopf units.}
\begin{tabular}{rccccccc}
\hline
\multicolumn{1}{c}{Transition} & Calc & Exp & Calc & Exp & Calc & Exp \\
\multicolumn{1}{c}{ } & \multicolumn{2}{c}{$A=134$} &
\multicolumn{2}{c}{$A=136$} & \multicolumn{2}{c}{$A=138$} \\
\hline
B(E2;$2^{+}_{1} \rightarrow 0^{+}_{1}$)\hspace{1.cm}SM & 3.5 & $>$ 0.024 & 7.0 & 
                                  9 (4) & 10.9 & 11.4 (3) \\ 
                               QRPA & 4.6 &    & 8.4 &   & 11.5 &  \\
B(E2;$4^{+}_{1} \rightarrow 2^{+}_{1}$)\hspace{1.cm}SM & 3.7 & 3.9 (4)  & 1.5 & 
                                  1.25 (6) & 1.5 & 0.286 (11) \\
B(E2;$6^{+}_{1} \rightarrow 4^{+}_{1}$)\hspace{1.cm}SM & 1.7 & 2.04 (5) & 0.3 & 
                                  0.0132 (5) & $< 10^{-4}$ & 0.053 (7) \\
B(E2;$10^{+}_{1} \rightarrow 8^{+}_{1}$)\hspace{.85cm}SM & & & & & 1.1 & 1.59 (22) \\
B(E3;$3^{-}_{1} \rightarrow 0^{+}_{1}$)\hspace{1.cm}SM & 1.6 $10^{-2}$ & & 0.2 & &
                                  1.6 & 0.133 (13) \\
                               QRPA & 12.6 & & 12.3 & & 12.7 & \\
\hline
\multicolumn{1}{c}{ } & \multicolumn{2}{c}{$A=140$} &
\multicolumn{2}{c}{$A=142$} & \multicolumn{2}{c}{$A=144$} \\
\hline
B(E2;$2^{+}_{1} \rightarrow 0^{+}_{1}$)\hspace{1.cm}SM & 13.4 & 16.6 (24) & 14.1 & 
        12.04 (18) & 15.1 & 11.5 (3)$^{\,{\rm a})}$ \\
                           QRPA & 14.8 &  & 14.2 &   & 11.2 &     \\
B(E2;$4^{+}_{1} \rightarrow 2^{+}_{1}$)\hspace{1.cm}SM & 0.1 & 0.137 (1) & & & & \\
B(E2;$6^{+}_{1} \rightarrow 4^{+}_{1}$)\hspace{1.cm}SM & 0.2  & 0.28 (6) & 0.05 & 
                           0.0179 (12) & 0.01 & 0.188 (10) \\
B(E2;$10^{+}_{1} \rightarrow 8^{+}_{1}$)\hspace{.85cm}SM & 0.5 & 0.46 (13) & & & & \\
B(E2;$0^{+}_{2} \rightarrow 2^{+}_{1}$)\hspace{1.cm}SM & 2.6  & 11.5 (9) & & & & \\
B(E3;$3^{-}_{1} \rightarrow 0^{+}_{1}$)\hspace{1.cm}SM & 2.9 & $> 5.6 (2)$ & 3.9 & 28.6 &
                             & 38 (3)$^{\,{\rm b})}$ \\
                            QRPA & 13.9 & & 16.2 & & 19.5 & \\  
\hline
\multicolumn{1}{l}{$^{{\rm a)}}$ {Ref.\ \cite{gvb90}}} & & & & & \\
\multicolumn{1}{l}{$^{{\rm b)}}$ {Ref.\ \cite{bbd89}}} & & & & & \\
\end{tabular}
\label{tab:even-e2-1}
\end{center}
\end{table}

In addition to the 0$^{+}$ ground state several excited states with 
$J = 0^{+}$ are known experimentally. The SM reproduces these states 
reasonably well. However, one $0^{+}_{2}\rightarrow 2^{+}_{1}$  E2 
transition in $^{140}$C  has been measured 
to 11.5~Wu and our  SM calculation fails in reproducing the strength of this
transition by a factor of more than four.
It is known that in shell model calculations with realistic interactions
E2 transitions like 
$2^{+}_{1} \rightarrow 0^{+}_{1}$ will be large whereas  
$0^{+}_{n} \rightarrow 2^{+}_{1}$ E2 transitons are small. This feature is 
due to an  enhanced quadrupole character of the effective interaction. Phenomenon like the present
one with a large $0^{+}_{2} \rightarrow 2^{+}_{1}$ transition indicates
the importance of other degrees of freedom in the wave functions, probably a significant
core excitation. For all the isotones more experimental information is therefore urgently
needed in order to identify the properties of these states in more detail. 

In the QRPA the first eigenstate
emerging from the diagonalization of the two-body interaction in
the $0^+$ coupled two-quasiparticle basis, is a spurious one \cite{bar60} and
related to the ground state. A convenient feature of the QRPA approach is that 
all the other eigenstates are free from such a spuriousity. 
The energy of the spurious state becomes zero leaving the rest of the 
excited $0^+$ states as physical ones. As seen from tables
\ref{tab:even-1} and \ref{tab:even-2} the $0^+$ 
QRPA reproduces the excited $0^{+}$ states  reasonably well in the middle of the shell
whereas it fails at the beginning and at the end, again due to the basic
BCS approximation used.

Of the negative parity states we have only calculated the 3$^{-}$ states. 
Such states may be generated through the $h_{11/2}$ negative parity orbit or 
through strong octupole core excitations.
The SM should reproduce states of the first type but fail for states of the second.
On the other hand we expect QRPA to describe strong octupole vibrational states since  
the single-particle orbits of the core are explicily included.  
Contrary to the $2^{+},\;4^{+}$ and $6^{+}$ states we find for the $3^{-}$ states clear 
differences between the two models. The QRPA produces always
a low-lying $3^{-}$ state not found in the SM
calculation. This may be a good candidate for 
a collective octupole state.
Unfortunately, we can not draw clear conclusions from the present
experimental data. In particular, the experimental information
on the E3($3^{-} \rightarrow 0^{+}$) transitions
as seen in Table~\ref{tab:even-e2-1}
is sparse and somewhat confusing.
In $^{142}$Nd and $^{144}$Sm a strong   E3($3^{-} \rightarrow 0^{+}$) is found 
experimentally in agreement with the QRPA prediction. On the other hand a 
very weak  E3($3^{-} \rightarrow 0^{+}$) transition  is measured in $^{136}$Xe and  
is more in line with the SM results. 
In the other cases of interest no experimental data are 
known. Thus as a temporary conclusion we have in  
Tables \ref{tab:even-1}, \ref{tab:even-2} and \ref{tab:even-3} for the SM 
results excluded
the lowest lying experimental $3^{-}$ 
when we identify our calculation with experiment.
The QRPA seems to reproduce all $3^{-}$ states with a question mark for the 
$3^{-}$ state in $^{136}$Xe.

Guided by our calculations we suggest the following interpretation for 
states given with several alternative experimental spin assignments:

The state in $^{136}$Xe at $2.126$ MeV is given the two alternative angular 
momentum values $J^{\pi} = 3^{+},4^{+}$. Our calculations 
give no $4^{+}$ state corresponding to this state, but the QRPA 
calculations predict a $3^{+}$ state with almost this energy.

In $^{138}$Ba there are two states where the angular momentum assignment is
tentative, and different values are considered. The first one is at $2.190$ 
MeV of excitation energy with $J=(1,2,3)$, and the other one is at $2.583$ MeV 
of excitation with $J^{\pi} = (1^{+},2^{+})$. Our calculations suggest that 
both states have $J^{\pi} = 2^{+}$.

In $^{140}$Ce there are two alternative assignments for the state of $2.516$
MeV excitation energy, $J^{\pi} = 3^{+}$ and $4^{+}$. Both our models 
predict a $4^{+}$ state near this energy. Our calculations do also give a 
$3^{+}$ state in the same energy region, but this most probably corresponds 
to the experimental $3^{+}_{1}$ state at $2.464$ MeV.

Of the negative parity states we have only calculated the 3$^{-}$ states. 
Here the deviation between experiment and the SM 
is significant, whereas the QRPA
reproduces the data reasonably well, except for $^{144}$Sm. 

The shell model is not able to reproduce the first $3^{-}$ state for any
of the isotones, while both QRPA approaches yield the lowest-lying $3^{-}$
states in nice agreement with experiment. The QRPA wave functions 
indicate that
these states are strongly collective, and that they are a mixture of both 
proton and 
neutron degrees of freedom. 
The philosophy of the effective theory is 
that the main components of the shell 
model wave functions have the origin  
within the model space ($P$-space components), 
and the rest ($Q$-space components) are included 
through the perturbation technique.
Not all degrees of freedom are taken into account this way, like for instance 
neutron core excitations. The above mentioned $3^{-}$ state, 
if assuming a strong mixture 
of neutron degrees of freedom, 
can therefore not be described within the frame of 
the shell model calculation. 
However, the $3^{-}$ states, which according to the 
QRPA are of two-quasiparticle character and consist of pure
proton excitations, are very well described by the shell model. 
This also indicates
that there is a minimal mixing between the different $3^{-}$ states. Our 
interpretation is then that the second experimental $3^{-}$ states contain
mainly two-quasiproton excitations with one exception. In $^{140}$Ce, 
it is the third experimental $3^{-}$ state which most probably corresponds to 
the two-quasiproton $3^{-}$ state. The $3^{-}$ state which appears in 
$^{140}$Ce at an excitation energy of $3.040$ MeV can neither be described 
within the QRPA nor the shell model. Similar low-lying $3^{-}$ states, which
can not be described within either of our models, are also observed in the 
other $N=82$ isotones. A possible explanation of the nature of these $3^{-}$ 
states might be that they are octupole deformed and possibly created by neutron
two-particle-two-hole excitations. These excitations activate the 
long-range part of the proton-neutron interaction and yield to deformation.

Calculations of the E3 transition $3^{-} \rightarrow 0^{+}$ can give 
information on whether our models succeed in describing the structure of the 
$3^{-}$ states or not. No B(E3;$3^{-}_{1} \rightarrow 0^{+}_{1}$) data are 
available for $^{134}$Te and $^{136}$Xe. In $^{138}$Ba there is a E3 
transition rate 
from the $3^{-}_{1}$ to the ground state which is measured to be 0.133 (13) 
W.u. There are uncertainties concerning the lifetime of the $3^{-}_{1}$ state
in $^{140}$Ce. An experiment by Grinberg et al. \cite{g93:E3} has 
given an upper limit for the lifetime $< 0.1$ ns, which 
corresponds to B(E3;$3^{-}_{1} \rightarrow 0^{+}_{1}$) $> 5.6 (2)$ W.u. In
$^{142}$Nd and $^{144}$Sm the B(E3;$3^{-}_{1} \rightarrow 0^{+}_{1}$) are
reported to be 28.6 W.u. and 38 (3) W.u., respectively.

The QRPA E3 transitions rates are fairly constant and vary from 12.6 W.u.
in $^{134}$Te to 19.6 W.u. in $^{144}$Sm. The size of the calculated B(E3) 
values  is a sign of collectivity 
and supports the picture we already have that the QRPA $3^{-}_{1}$ states 
are due to excitations of the core. For the same reasons as in the
E2 calculations, no effective charge is here used.

The SM B(E3) values increase with increasing number of valence particles.
An effective charge of $e_{\rm eff} = 2.7e$ will be required in order to 
reproduce the B(E3;$3^{-}_{1} \rightarrow 0^{+}_{1}$) in $^{142}$Nd. 
Due to the strong dependence of the number of active particles it is 
likely to believe that the $3^{-}_{1}$ SM states are results of pure 
excitations of the valence protons.

Before we can draw further conclusions more thorough measurements of 
the $3^{-}$ lifetime in $^{140}$Ce is needed, and we will also encourage the 
experimentalists to seek more information about the $3^{-}$ states in
the N=82 isotones lighter than $^{140}$Ce.

\subsection{Generalized seniority}

The  proton occupation number 
for each $(lj)$  value is determined within the shell model and compared with 
experiment and the BCS occupations underlying the QRPA calculation in Table\ 
\ref{tab:occupation}. For both the shell model and the BCS the calculated 
ground states have too large components of $d_{5/2}$, but the sum of the 
$g_{7/2}$ and $d_{5/2}$ occupation numbers are close to the experimental 
values. The proton occupation numbers, calculated for the ground
state of $^{146}$Gd, show that on the average $12$ out of $14$ valence 
particles are occupying the $0g_{7/2}$ and $1d_{5/2}$ orbitals. As a general
observation one can see from Table \ref{tab:occupation} that there is, in 
general, a close correspondence between the shell model occupations and the
BCS occupations.

\begin{table}[htbp]
\begin{center}
\caption{Experimental (Ref.\ \protect \cite{wna71}) and calculated proton occupation 
numbers for the ground states of the even-mass $N=82$ isotones.}
\begin{tabular}{lcccccc}
\hline
s.p. orbitals & $g_{7/2}$ & $d_{5/2}$ & $h_{11/2}$ & $d_{3/2}$ & $s_{1/2}$ & 
            $g_{7/2} + d_{5/2}$ \\
\hline
$^{134}$Te & & & & & & \\
SM         & 1.56   & 0.25   & 0.13   & 0.05   & 0.02   & 1.81 \\
BCS        & 1.60   & 0.30   & 0.20   & 0.04   & 0.02   & 1.90 \\
$^{136}$Xe & & & & & & \\
Exp        & 3.5(4) & 0.5(2) & 0.0(7) & 0.0(2) & 0.0(2) & 4.0(6) \\
SM         & 2.79   & 0.76   & 0.28   & 0.12   & 0.04   & 3.55 \\ 
BCS        & 3.02   & 0.63   & 0.46   & 0.09   & 0.03   & 3.65 \\
$^{138}$Ba & & & & & & \\
Exp        & 4.3(4) & 0.7(3) & 1.0(8) & 0.0(2) & 0.0(2) & 5.0(7) \\
SM         & 3.69   & 1.57   & 0.45   & 0.22   & 0.07   & 5.26 \\
BCS        & 4.20   & 1.27   & 0.59   & 0.15   & 0.05   & 5.47 \\
$^{140}$Ce & & & & & & \\
Exp        & 5.6(3) & 1.8(2) & 0.6(4) & 0.0(2) & 0.0(2) & 7.4(5) \\
SM         & 4.44   & 2.50   & 0.62   & 0.34   & 0.16   & 6.94 \\
BCS        & 4.55   & 2.55   & 0.80   & 0.32   & 0.09   & 7.10 \\
$^{142}$Nd & & & & & & \\
Exp        & 5.7(3) & 2.6(3) & 1.3(5) & 0.2(1) & 0.2(1) & 8.3(6) \\
SM         & 5.17   & 3.43   & 0.78   & 0.46   & 0.16   & 8.60 \\
BCS        & 5.62   & 3.32   & 0.86   & 0.36   & 0.10   & 8.94 \\
$^{144}$Sm & & & & & & \\
Exp        & 6.3(2) & 3.6(2) & 1.6(3) & 0.3(1) & 0.2(1) & 9.9(4) \\
SM         & 5.90   & 4.30   & 0.96   & 0.61   & 0.23   & 10.20  \\
BCS        & 6.63   & 4.29   & 0.82   & 0.32   & 0.12   & 10.92 \\
$^{146}$Gd & & & & & & \\
SM         & 6.72   &   4.96 & 1.15   & 0.82   & 0.35   & 11.68 \\
BCS        & 7.28   & 5.14   & 0.98   & 0.43   & 0.30   & 12.42 \\
\hline
\end{tabular}
\label{tab:occupation}
\end{center}
\end{table}

The results from Table \ref{tab:occupation} indicate that
pairing effects are likely to be important. In order to test
whether generalized seniority is conserved, we construct the pair correlation 
operator for creating a generalized seniority $v=0$ pair
\begin{equation}
     S^{\dagger} = \sum_{j} \frac{1}{\sqrt{2j + 1}} C_{j} \sum_{m \geq 0}
                   (-1)^{j-m} a^{\dagger}_{jm} a^{\dagger}_{j-m},
\end{equation}
where the coefficients $C_{j}$ are obtained from the ground state of
$^{134}$Te. Similarly the generalized seniority $v=2$ operator takes the
form
\begin{equation}
    D^{\dagger}_{J=2,M=0} = \sum_{j \leq j',m\geq 0} (1+\delta_{j,j'})^{-1/2}
                            \beta_{j,j'}\langle jmj'-m|20\rangle 
                            a^{\dagger}_{jm} a^{\dagger}_{j'-m},
\end{equation}
where the coefficients $\beta_{j,j'}$ are obtained from the first excited 
$2^{+}$ state of $^{134}$Te. 

With our shell model wave functions we evaluate the squared overlaps
$|\langle A; J^{\pi}| S^{\dagger} | A-2; J^{\pi} \rangle |^{2}$ and
$|\langle A; 2^{+}_{1}| D^{\dagger}_{J=2, M=0}| A-2; 0^{+}_{1} \rangle |^{2}$.
The results are given in Tables\ \ref{tab:even-sen-1} 
and \ref{tab:even-sen-2}.

\begin{table}[htbp]
\begin{center}
\caption{Generalized seniority overlap 
$|\langle A;J_{f}|S^{\dagger}|A-2;J_{i}\rangle |^{2}$ where $J_{i} = J_{f}$
and the generalized seniority overlap 
$|\langle A; J_{f}|D^{\dagger}|A-2;J_{i}\rangle |^{2}$ with $J_{i} = 0$ and
$J_{f} = 2$ for the low-lying eigenstates of $^{134}$Te - $^{140}$Ce.}
\begin{tabular}{cccc}
\hline
$A-2 \rightarrow A$ & $^{134}$Te $\rightarrow ^{136}$Xe & 
$^{136}$Xe $\rightarrow ^{138}$Ba & $^{138}$Ba $\rightarrow ^{140}$Ce \\
$J_{i}^{\pi} \rightarrow J_{f}^{\pi}$  & $A=136$ & $138$ & $140$ \\
\hline
$0^{+}_{1}(v=0) \rightarrow 0^{+}_{1}(v=0)$ & 0.982   & 0.962   & 0.957   \\
$2^{+}_{1}(v=2) \rightarrow 2^{+}_{1}(v=2)$ & 0.939   & 0.899   & 0.931   \\
$4^{+}_{1}(v=2) \rightarrow 4^{+}_{1}(v=2)$ & 0.948   & 0.967   & 0.928   \\
$6^{+}_{1}(v=2) \rightarrow 6^{+}_{1}(v=2)$ & 0.896   & 0.753   & 0.913   \\
$0^{+}_{2}(v=0) \rightarrow 0^{+}_{2}(v=0)$ & 0.853   & 0.871   & 0.761   \\
$0^{+}_{1}(v=0) \rightarrow 2^{+}_{1}(v=2)$ & 0.939   & 0.761   & 0.613   \\
\hline
\end{tabular}
\label{tab:even-sen-1}
\end{center}
\end{table}

\begin{table}[htbp]
\begin{center}
\caption{Generalized seniority overlap 
$|\langle A;J_{f}|S^{\dagger}|A-2;J_{i}\rangle |^{2}$ where $J_{i} = J_{f}$
and the generalized seniority overlap 
$|\langle A; J_{f}|D^{\dagger}|A-2;J_{i}\rangle |^{2}$ with $J_{i} = 0$ and
$J_{f} = 2$ for the low-lying eigenstates of $^{140}$Ce - $^{144}$Sm.}
\begin{tabular}{ccc}
\hline
$A-2 \rightarrow A$ & $^{140}$Ce $\rightarrow ^{142}$Nd & 
$^{142}$Nd $\rightarrow ^{144}$Sm \\
$J_{i}^{\pi} \rightarrow J_{f}^{\pi}$  & $A=142$ & $144$ \\
\hline
$0^{+}_{1}(v=0) \rightarrow 0^{+}_{1}(v=0)$    & 0.958   & 0.959   \\
$2^{+}_{1}(v=2) \rightarrow 2^{+}_{1}(v=2)$    & 0.911   & 0.889   \\
$4^{+}_{1}(v=2) \rightarrow 4^{+}_{1}(v=2)$    & 0.888   & 0.824   \\
$6^{+}_{1}(v=2) \rightarrow 6^{+}_{1}(v=2)$    & 0.915   & 0.866   \\
$0^{+}_{2}(v=0) \rightarrow 0^{+}_{2}(v=0)$    & 0.796   & 0.781   \\
$0^{+}_{1}(v=0) \rightarrow 2^{+}_{1}(v=2)$    & 0.513   & 0.475   \\
\hline
\end{tabular}
\label{tab:even-sen-2}
\end{center}
\end{table}

The ground state can be well described within a generalized seniority scheme.
More than $95\%$ of the wave function for the $(A)$ system is given as the 
$(A-2)$ system plus a seniority-zero pair. Also for the $2^{+}_{1}$, 
$4^{+}_{1}$ and $6^{+}_{1}$ states, generalized seniority is an 
approximately conserved quantum number. The overlap is in general $85-95\%$.

Andreozzi {\sl et al.} \cite{acgp90} pointed out the importance of 
seniority-zero components in the structure of the $0^{+}_{2}$ states. 
They managed to reproduce nicely the $0^{+}_{2}$ energy levels 
within their pairing model, but no E2 transitions were calculated which 
would give a more sensitive test of the wave functions.

In $^{134}$Te the $0^{+}_{2}$ state is not yet observed, but for the other 
nuclei, $^{136}$Xe - $^{144}$Sm, data is now available. 
As mentioned before we are not able to give such a good description of 
these states. Only $75 - 85\%$ of our states are given as the $0^{+}_{2}$ 
state in the $(A-2)$ system plus a seniority-zero pair.

\section{Conclusions}

In this work a comprehensive study of the $N=82$ isotones has been 
presented. We have performed a large-basis shell model calculation with 
$2 - 12$ valence protons outside  the doubly magic nucleus $^{132}$Sn. The 
main ingredient in the shell model calculation is a realistic, 
microscopic two-body effective interaction derived from a modern 
meson-exchange NN potential using many-body perturbation theory. 
The philosophy behind the perturbative approach is to include degrees
of freedom not accounted for in the model space defining the effective
interaction and the shell-model problem. The hope is then that
degrees of freedom not included in the shell-model
space can be accounted for by such renormalized effective
interactions. Differences which may arise between the shell-model
approach and the QRPA, where a larger single-particle basis is employed,
may then shed light on the strengths and the limitations of the two approaches.
As seen from our results, 
the low-lying states of the even $N=82$ isotones are, in general, very well
described, both by the shell model and the QRPA. The fact that the QRPA allows
for proton and neutron core excitations, gives this model the flexibility to
describe some collective excitations 
which are out of reach of the present shell 
model description. An example are the first excited $3^{-}$ states, which can 
not be described by the shell model. The QRPA indicates that all these 
$3^{-}_{1}$ states are strongly collective, and that also non-negligible 
components of neutron excitations contribute. However, the somewhat 
higher-lying $3^{-}$ states, implied by the QRPA to be of two-quasiproton
character, are nicely described by the shell model. There are however other
$3^{-}$ states that can not be described 
by any of the models, i.e. in $^{140}$Ce.

The enhanced E2 transitions between the first excited $2^{+}$ state and the 
$0^{+}$ ground state, indicate that the $2^{+}_{1}$ states are of vibrational
collective nature. With both models we obtain $B(E2;2^{+}_{1} \rightarrow 
0^{+}_{1})$ values in good agreement with experiment. This is an indication 
that the calculated wave functions for these states contain the relevant 
components. In general, the shell model does also give E2 transitions between 
the other yrast states in good agreement with data.

The aim of this work was to test a new effective interaction for
protons outside the closed $^{132}$Sn core. In view of the fact that 
this is a truly  microscopic calculation, with very few 
parameters, the agreement with data is remarkably good.
Our effective interaction seems to be more successful in the region of 
medium heavy nuclei than for light nuclei (oxygen and calsium regions). 
The individual degrees of freedom may be more important for light nuclei than 
for heavier systems, and therefore the results more sensitive to 
the fine details of the effective interaction.

This work has been supported by the NorFA (Nordic Academy for Advanced Study).
The work of M.H.J.\ has been supported
by  the Instituto Trentino di Cultura, Italy, and the Research 
Council of Norway (NFR). The calculations have been 
carried out in the IBM cluster at the University of Oslo. Support for this
from the NFR is acknowledged.

\end{document}